# The mass of the black hole in LMC X-3

A.K.F. Val-Baker[*] A.J. Norton[*] and I. Negueruela[†]

[*]*Department of Physics and Astronomy, The Open University, Milton Keynes MK7 6AA, U.K*
[†]*DFISTS, Universidad de Alicante, Apartado de Correos 99, E03080 Alicante, Spain*

**Abstract**
New high resolution, optical spectroscopy of the high mass X-ray binary LMC X-3, shows the spectral type of the donor star changes with phase due to irradiation by the X-ray source. We find the spectral type is likely to be B5V, and only appears as B3V when viewing the heated side of the donor. Combining our measurements with those previously published, and taking into account the effects of X-ray irradiation, results in a value for the donor star radial velocity semi-amplitude of $K_o = 256.7 \pm 4.9 \, \text{km/s}$. We find the mass of the black hole lies in the range $9.5 M_\odot \leq M_x \leq 13.6 M_\odot$.

**Keywords:** stars: individual: LMC X-3 - stars: fundamental parameters - binaries: close

**PACS**: 97.10.Nf, 97.60.Lf, 97.80.-d

## INTRODUCTION

LMC X-3 is a non-eclipsing massive X-ray binary in the Large Magellanic Cloud. Originally detected as an X-ray source by Leong et al. (1971), the optical counterpart was identified as a faint (V ~ 16.7 – 17.5) OB star by Warren & Penfold (1975) and the spectral type was subsequently classified as B3V (Cowley, et al. 1983). However, this spectral type remains uncertain, due to the effects of irradiation by the X-ray source, which in turn leads to uncertainties in the mass determinations of the compact object.

## OBSERVATIONS AND DATA REDUCTION

The observations presented here were obtained with UVES on the ESO VLT. Five spectra were taken between December 2004 – March 2005. The blue spectra used here span a wavelength range of $\sim 3780 \text{Å} - 4980 \text{Å}$ with a resolution of $0.02 \text{Å}$ per pixel. All spectra were reduced using standard IRAF routines and extracted using optimal extraction procedures. The extracted spectra were then continuum fitted using DIPSO. Orbital phases corresponding to each spectrum were calculated using the ephemeris from van der Klis et al. (1985), which gives the time of the $N^{th}$ superior conjunction of the X-ray source as:

$$T_N / \text{HJD} = 2445278.005(\pm 0.011) + 1.70479(\pm 0.00004) N$$

A barycentric correction was applied to the HJD, before the phases were calculated.

# SPECTRAL CLASSIFICATION

There are no X-ray pulsations with which to probe the motion of the compact object and hence determine its radial velocity (RV) semi-amplitude $K_x$. Therefore, the mass of the optical companion has to be deduced from its spectral type, which can lead to huge uncertainties if the spectral type is not known accurately. The optical counterpart to LMC X-3 is currently classified as a B3V star, but due to X-ray irradiation of the optical companion, the spectral type appears to change with phase. We compared our LMC X-3 spectra with standard star spectra from `The OB Spectral Classification Atlas' (Walborn and Fitzpatrick 1990), `The standard star catalogue' (Jacoby, Hunter and Christian 1984) and archive data from the ESO VLT. The spectra obtained at phase 0.22, 0.34, 0.63 and 0.64, when the heated face of the optical companion is in the line of sight, show relatively strong helium and hydrogen lines with little variance between each spectrum. As the spectrum obtained at phase 0.22 has the best signal to noise we used this to determine the spectral type of the star when seeing its heated side. This was found to be most similar to B3V (Fig. 1a). However, the spectrum obtained at phase 0.86 shows much weaker helium lines (Fig. 2a) and narrower and stronger hydrogen lines (Fig. 2b). This is to be expected, as at this phase the optical companion is nearly in the line of sight and the black hole is almost out of view, so we are seeing the cooler, unheated side of the companion star. The spectrum at this phase was found to be most similar to B5V (Fig. 1b).

From Kurucz's stellar atmosphere models (Kurucz 1979, 1993), the approximate effective temperature of a B3V star is 18700 K, and of a B5V star is 15400 K. Therefore, the star is being heated as a result of X-ray irradiation by ~ 3300 K.

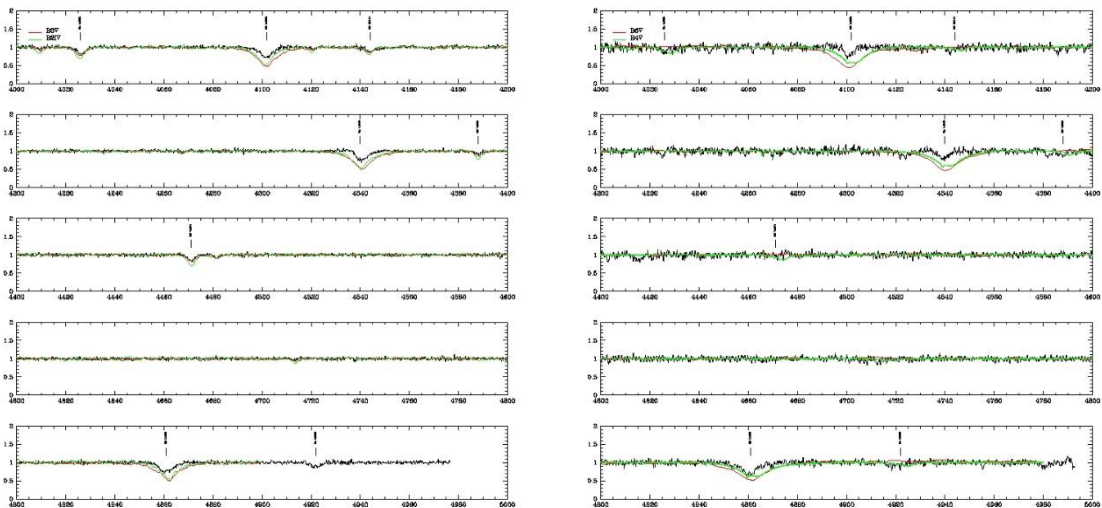

Figure 1: (a) The continuum normalized optical spectrum of LMC X-3 at phase 0.22, shown with B2IV and B3V standard star spectra for comparison. (b) The continuum normalized optical spectrum of LMC X-3 at phase 0.86, shown with B4V and B6V standard star spectra for comparison}

# SYSTEM PARAMETERS FROM THE RV CURVE

Gaussian fits to the HeI $4026\text{Å}$ and HeI $4713\text{Å}$ absorption lines were performed on each of the five spectra. The heliocentric corrected velocities were then plotted against phase, along with the RV measurements of Cowley et al. (1983). In order to achieve agreement, a phase shift of

0.05 was required, but this is within the extrapolated uncertainty of the van der Klis et al. (1985) ephemeris, so is not unexpected. The resulting radial velocities were fitted with a sinusoid, allowing the mean and amplitude as free parameters. The semi-amplitude of the RV curve is $K_o = 242.4 \pm 4.3\,\text{km/s}$ and the systemic velocity is $\gamma = 301.7 \pm 3.5\,\text{km/s}$. The mass of the X-ray source is given by $M_x = K_o^2 P(1+1/q)^2 / 2\pi G \sin^3 i$ where $q(=M_x/M_o)$ is the mass ratio and $i$ is the inclination of the orbital plane to the line of sight. Since the system is not eclipsing we can assume that $i < 70°$, and from the measured limit on rotational broadening we know that $i > 50°$. Taking the mass of the optical companion to be that of a B5V star (i.e. $M_o \sim 5.9 M_\odot$), gives a mass range of the X-ray source of $8.6 M_\odot \leq M_x \leq 12.3 M_\odot$.

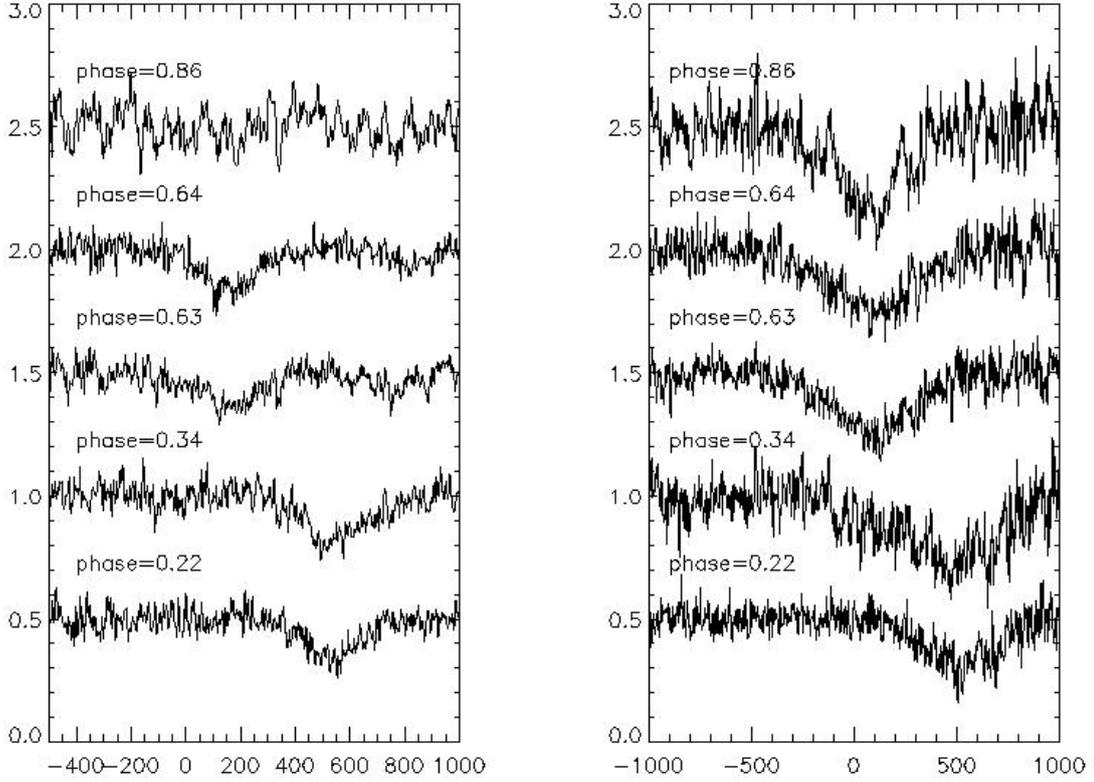

Figure 2: (a) The HeI $4471\text{Å}$ absorption line shown at different phases. (b) The Balmer H10 absorption line shown at different phases.

## X-RAY HEATING CORRECTIONS

X-ray irradiation of the optical companion by the X-ray source significantly alters the observed RV amplitude and so distorts the inferred compact star mass. Due to the heating and temperature dependence of the spectral lines, the shift in line centre may vary significantly with the intensity of irradiation, and can therefore have one of two different effects on the observed RV curve of the optical companion. If the X-ray spectrum is soft, the X-rays will not be able to penetrate into the continuum forming layers and instead are absorbed at the surface of the companion star, so infilling the absorption lines. As a result, the effective centre of light of the companion star integrated over the entire stellar disk is shifted *away* from the centre of mass of the system, such that the observed velocity amplitude is *greater* than the true Keplerian RV amplitude. Conversely, if the X-ray spectrum is hard the X-rays are either absorbed deep in the atmosphere of the companion star or directly reflected, causing the absorption lines from the heated atmosphere of the star to be stronger than they would otherwise be. As a result, the effective centre of light of the companion star integrated over the entire stellar disk is shifted *towards* the centre of mass of the system, and this *reduces* the observed velocity amplitude to be less than the true Keplerian RV amplitude.

LMC X-3 is known to exist in both the soft and hard X-ray states, but is usually found in the soft state. Previous authors (e.g. Soria, et al. 2001) have therefore assumed that the X-ray irradiation will lead to exaggerated RV amplitudes. However, our spectra show that the HeI absorption lines are stronger from the heated side of the companion star, which suggests that we are seeing the effects of hard X-rays rather than soft X-rays. Since it is these lines that are used to generate the RV curve, the true RV curve must have larger amplitudes than those observed.

To correct for the heating effects, we followed the procedure in our earlier paper (Val Baker, Norton and Quaintrell 2005) and ran models using *LIGHT2* (Hill 1988), a sophisticated light-curve synthesis program. We set the radius and polar temperature of the object representing the black hole so as to produce a blackbody luminosity equivalent to that needed to increase the temperature of the companion star by 3300 K. The code was required to run through three iterations before convergence. The semi-amplitude of the corrected RV curve (Fig. 3) is $K_o = 256.7 \pm 4.9$ km/s and the systematic velocity is $\gamma = 300.0 \pm 4.1$ km/s. Therefore, the mass of the X-ray source lies in the range $9.5 M_\odot \leq M_x \leq 13.6 M_\odot$. This mass is comparable to those found for the black holes in the HMXB Cyg X-1 and LMC X-1 and many of the LMXB black hole candidates.

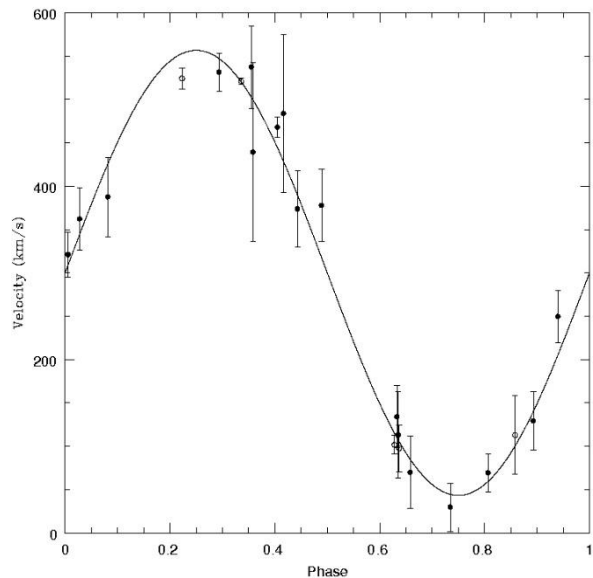

Figure 3: Heating corrected RV curve for LMC X-3, when $i = 70°$. Our data are shown as open circles and those of Cowley et al. (1983) are shown as filled circles.